\documentclass[apjl]{emulateapj}

\begin{document} 
\title{UCD Candidates in the Hydra Cluster}
\submitted{Accepted for publication in ApJ Letters}
\author{Elizabeth M. H. Wehner\altaffilmark{1,2} and William E. Harris\altaffilmark{1,3}}
\altaffiltext{1}{Department of Physics \& Astronomy, McMaster University, Hamilton, ON L8S 4M1, Canada}
\altaffiltext{2}{wehnere@physics.mcmaster.ca} 
\altaffiltext{3}{harris@physics.mcmaster.ca} 

\shorttitle{UCD Candidates in the Hydra Cluster}
\shortauthors{Wehner \& Harris}

\begin{abstract}

	NGC~3311, the giant cD galaxy in the Hydra cluster (A1060), 
has one of the largest globular cluster systems known. We describe new Gemini GMOS $(g',i')$ photometry of the NGC~3311 field which reveals that the {\sl red, metal-rich} side of its globular cluster
population extends smoothly upward into the mass range associated with the new class of
Ultra-Compact Dwarfs (UCDs). We identify 29 UCD candidates with estimated masses $> 6 \times
10^6 M_{\odot}$ and discuss their characteristics. This UCD-like sequence is the most well
defined one yet seen, and reinforces current ideas that the high-mass end of the globular
cluster sequence merges continuously into the UCD sequence, which connects in turn to the E
galaxy structural sequence.

\end{abstract}


\section{Introduction} 
\label{intro}

Ultra-Compact Dwarfs (UCDs) are a recently discovered type of old stellar system, 
lying between the classic globular clusters and dwarf elliptical galaxies 
in luminosity and scale size.  Initially discovered in the 
Fornax cluster \citep{hilker98,hilker99,dw00,phillipps01,mieske02}, UCDs 
and UCD candidates have since been discovered in Abell 1689 \citep{mieske04b} and
the Virgo Cluster \citep{hasegan05,jones06,evsti07}.  
Transitional objects on the lower-mass end of 
the UCD range that connect closely with the high-mass end of 
the globular cluster sequence, have also been found in NGC 5128 \citep{rejkuba07, mieske07}.   

Because UCDs have scale radii typically $r_{eff} \lesssim 30$ pc (not much different from extended, luminous GCs, or the nuclei of dE,Ns), they are extremely hard to find by morphology or image structure alone at galaxy distances much beyond the Virgo or Fornax clusters.  Thus as yet, we know of very few UCDs.  To understand what sorts of galaxy environments are most likely to produce them, we need to use a wider variety of search methods in many more locations.  
One such method is to employ their \textit{photometric} 
similarity to the most luminous GCs and dE nuclei.  If large numbers of UCDs 
are present in a cluster of galaxies, then they might show up 
as high-luminosity extensions of the normal, bimodal GC sequences that we 
conventionally find around giant galaxies \citep{harris06, peng06}.  These 
candidates can then be followed up on via spectroscopy to determine their 
cluster memberships properties, such as metallicity, mass and age \citep[e.g.][]{evsti07}.

The cD galaxy NGC~3311 is the centrally dominant 
elliptical in the nearby Hydra cluster (Abell 1060) at
d=54 Mpc ($H_0 = 73$ km/s, $\Omega_M = 0.27$, $\Omega_{\Lambda} = 0.73$ from NED) 
and is an excellent candidate for UCD-based searches of this type.
The Hydra cluster ($v = 3777$ km/s) with 157 galaxy members listed by
\citet{struble99} is perhaps twice as rich as the Fornax Cluster ($v = 1379$ km/s) in which the largest numbers of UCDs have been detected thus far. Previous
photometric studies of NGC 3311 show that it contains one of the
richest globular cluster (GC) systems in the local universe
\citep{sw76, harris83, secker95, mclaughlin95, brodie00}, making it an excellent target to
search for unusually massive clusters and stripped dE nuclei.

As part of a new imaging program to investigate globular cluster systems (GCSs) around cD
galaxies, we obtained deep $(g',i')$ photometry of NGC~3311 to investigate the nature of the
new ``mass/metallicity relation" recently discovered to affect the 
metal-poor GC sequence
\citep{harris06, strader06, mieske06}. Our results have revealed 
an extension of the red, metal-rich
branch of the globular cluster system up to unusually high luminosities ($-10 > M_{g'} >
-12$), into the UCD regime. In \S~\ref{obs_red} we present our observations and data
reduction. In \S~\ref{results} we examine the radial distributions and masses of our
candidate UCDs, and we discuss the implications of our results in \S~\ref{discuss}.

\section{Observations and Reductions}
\label{obs_red}

We obtained deep ($g'$,$i'$) images of NGC~3311 using the GMOS imager on Gemini South, which
has a $5.5\arcmin \times 5.5\arcmin$ field of view (FOV) and a scale of
0.146$\arcsec/pix$ after $2\times2$ binning. Data were taken on the nights of February 8 and
March 23, 2006, under dark, photometric conditions, with an average seeing of
$0.5\arcsec$. The total integration time in each of ($g'$,$i'$) was 3900s. The data were
reduced with the GEMINI package in IRAF\footnote{IRAF is distributed by the National Optical
Astronomy Observatories, which are operated by the Association of Universities for Research
in Astronomy, Inc., under cooperative agreement with the National Science Foundation.},
calibrated with Landolt standard stars \citep{landolt92}, and transformed to ($g',i'$) with
the equations of \citet{fukugita96}\footnote{A complete set of the \citet{landolt92} standard
stars transformed from the Johnson-Cousins system into the Sloan filter set has been compiled
by the authors and is available online at
http://www.elizabethwehner.com/astro/sloan.html}. For the photometric calibration, we
used standard stars (with a very limited range in airmass) to measure a zero point for single
($g',i'$) exposures taken at the same airmass as our NGC 3311 images. 
The small
amount of fringing in $i'$ was successfully removed with a calibration fringe
frame from the Gemini archives. Our photometry reached limiting magnitudes (50\% detection
completeness) of $g'(lim) = 26.7$ and $i'(lim) = 26.2$, deep enough to reach near the GC
luminosity function turnover point.

\section{Photometric Results}
\label{results}

Once the final image and calibration were obtained, we used the stand-alone version of DAOPHOT (daophot4) to 
obtain photometric measurements in $g'$ and $i'$ of each object in our $5.5' $
FOV.  In total, we detected 8108 starlike objects, the vast 
majority of which clearly belong to the globular cluster population.  
The color-magnitude diagram (CMD) is shown in Figure 1.\footnote{Our FOV contains the 
Hydra gE NGC~3309, which has been found to contribute $\lesssim$ 10\%  
of the GC population in the field  \citep{mclaughlin95,brodie00}. 
Nevertheless, we exclude only those objects in the  
inner $r = 120$ pixels around the centers of both giant galaxies.}

%
%
\begin{figure}
\plotone{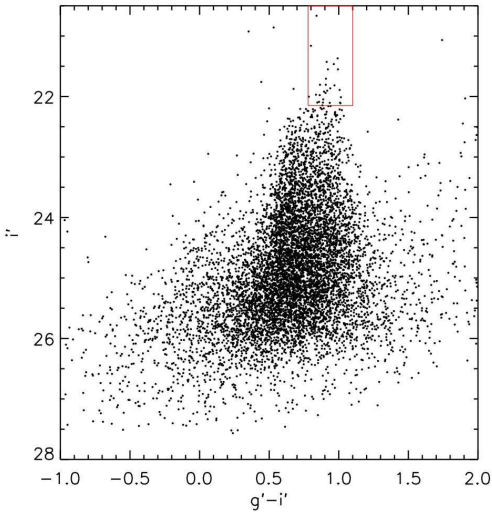}
\epsscale{0.5}
\figurenum{1} 
\caption{ Color-magnitude Diagram for the globular clusters around
NGC~3311.  Note that the color index plotted here is the
{\sl dereddened} value $(g'-i')_0$.  The box (outlined in red in the 
online version) indicates the location in color-magnitude space of the UCD candidates. }
\label{fig1} 
\end{figure}

The CMD shows the GC population in the expected color range 
$0.4 \lesssim g'-i' \lesssim 1.2$ as well as field contamination 
at faint levels on both the redder and bluer sides.  
Although this CMD is interesting in several ways (to be discussed in our 
upcoming Paper II), perhaps the most unusual feature can be found at the 
brightest magnitudes where we note the presence of an extension up to very 
high luminosities on the red (metal-richer) side of the GC population.  This 
distribution is unlike any we have seen before in other giant E 
galaxies \citep[e.g.][]{harris06}.  We find 29 objects brighter than 
$i' \leq 22.15$, the point where the blue side of the GC distribution
reaches its top end.  Intriguingly, it is only the red branch of the GC population
that extends toward still higher luminosities.  
At $M_{i'} \lesssim -11.7$, these are very luminous GCs.  
Few such objects appear even in the composite sample of many 
thousands of GCs in eight cD galaxies studied by \citet{harris06}, and easily 
reach up to the range occupied by UCDs and dE nuclei.  Because of their 
connection with the red GC sequence, these UCD candidates could also be 
categorized as dwarf-globular transition objects (DGTOs), 
a term coined by \citet{hasegan05} to indicate the difficulty in truly 
distinguishing compact dwarfs from massive globular clusters.

How many of these objects are real?  One concern may be blending, 
i.e. two ordinary globular clusters aligned along the line of sight, 
boosting the observed magnitude up to $\sim0.75$ magnitudes.  However, 
using the equation for the number of expected blends on a frame 
from \citet{harris07}, we find that for GCs brighter than $i' \simeq 23$, 
we expect only 0.47 blends across our entire frame.  
If instead these objects were foreground stars, we would expect 
them to be more evenly distributed in color and randomly distributed 
across the frame (as we show is not the case in Figures 2 and 3).

We also note that four objects fall blueward of the clear sequence seen in Fig.~1.  
Although perhaps not part of this sequence, these may be UCDs as well, and so we 
include them.  One of these objects (number 13 in Table 1) falls near 
the top of the blue sequence, although we cannot determine if this is actually 
a member/extension of the blue sequence, or merely a coincidental overlap. 
The remaining objects, those in the red sequence, extend up to $M_{i'} = -12.4$ ($i' = 21.4$).  

One of our first goals is to establish whether or not these objects are associated with
NGC~3311.   Figure 2 shows the location of each 
UCD candidate.  Close visual inspection of 
each object reveals them to be (apparently) stellar in nature, not obviously
background galaxies.  Furthermore, since they appear to be distributed across 
the field of view rather than clumped, we can rule out contamination from 
a distant background galaxy cluster. 

%
%
\begin{figure}
\figurenum{2} 
\plotone{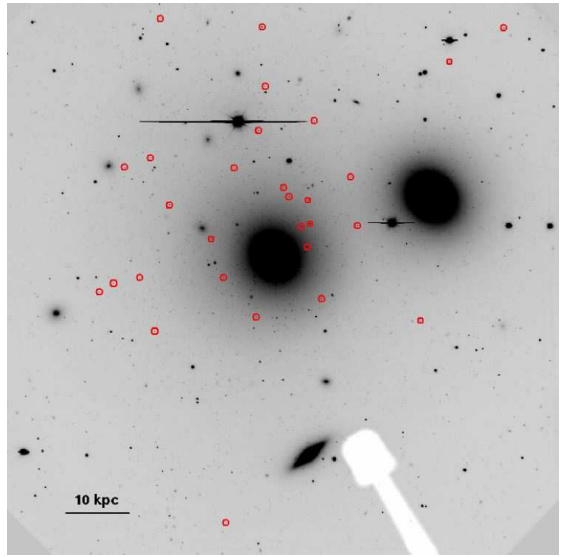}
\caption{$i'$-band GMOS field of NGC~3311 (center) and NGC~3309 (upper right).  Locations of UCD candidates are marked by circles.  The field size shown is $5.5\arcmin$ across.  The shadow of the guide probe is seen at the bottom.  North is up, and East is to the left.} 
\label{fig2} 
\end{figure}

But how do these objects relate to NGC~3311? Figure 3 shows the cumulative
radial distribution for the UCD candidates, relative to the center of NGC~3311,
as well as the cumulative distribution for the GC population as a whole. 
(In order to plot the GCS radial distribution, we used only
globular clusters on the side of the field opposite to NGC~3309.) 
A Kolmogorov-Smirnov two-sample test on these two radial distributions 
shows them to be different at more than 99 percent confidence. The
difference is in the sense (see Figure 3) that the UCDs are {\sl at least} as centrally
concentrated as the GCs; in other words, they are more likely to be connected with the GCS
rather than the more extended Hydra potential well as a whole.

%
%
\begin{figure} 
\plotone{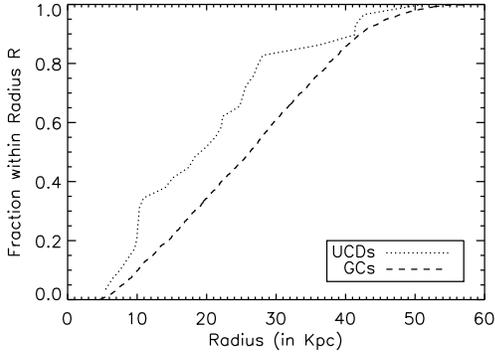} 
\figurenum{3} 
\caption{Radial distribution of UCD candidates (dotted line) 
and globular clusters (dashed line) relative to the center of NGC~3311.  
The distributions are plotted in cumulative form, as the fraction of
the total population lying within projected radius R.}
\label{fig3} 
\end{figure}

Next we estimate the mass of each UCD candidate (hereafter UCDs), for which the choice of $M/L$ is critical.  While globular clusters typically have $(M/L)_V \sim 1-3$, UCDs have $M/L$ as high as 6-9 $M_{\odot}/L_{\odot}$ \citep{hasegan05,evsti07}, although simulations by \citet{fk06} suggest that tidal interactions with the centers of their host galaxies \textit{may} disturb a UCD from its virial equilibrium, thereby leading to an overestimation of its $M/L$ ratio.  \citet{hilker07} found that 5 UCDs in Fornax ranged from 3-5 $M_{\odot}/L_{\odot}$, exactly the same $(M/L)_V$ range found for Virgo UCDs \citep{evsti07}.  If we assume, rather, that our objects are more similar to high-mass globular clusters, then we can look to simple stellar population models for predicted $M/L$ ratios.  \citet{bc03} models for old stellar populations ($t \sim 12$ Gyr) indicate that $(M/L)_V$ can range from 1-6 $M_{\odot}/L_{\odot}$, increasing systematically with metallicity.  The red peak of the GCS falls 
near $(V-I)_0 \sim 1.1$, or $[Fe/H] \sim -0.5$, which in 
the Bruzual \& Charlot models corresponds to $M/L \sim 2$.  The average $M/L$ ratio in \citet{rejkuba07}'s newly discovered relationship between $M/L$ and mass for UCDs is $\sim 3$.  
Since this most closely represents the undefined nature of our objects, 
and is a midpoint between the aforementioned low ($M/L \sim 1$) and high ($M/L \sim 5$) ends, 
we adopt a compromise $(M/L)_V = 3$ for our UCD candidates. 

We used the $g'$ magnitudes to calculate a total luminosity for 
each UCD in our sample.  The solar luminosity $M_{g'}{\odot} = 5.06$ 
was adopted from \citet{yasuda01}.  The $g'$-band was chosen over 
the $i'$-band for calculating masses in order to most closely match 
the expected $M/L$ ratio, which is well documented for both the Johnson $B$ and $V$ 
bands but not so extensively in either the infrared or the Sloan filter system.  
In order to calculate absolute magnitudes, we adopted a distance modulus 
of $(m-M)_0 = 33.68$ 
and a reddening of $E_{(g'-i')} = 0.158$ (from NED).
The raw $i'$ magnitudes, the $(g'-i')_0$ 
dereddened colors, and the estimated masses for each object are 
listed in Table 1, along with the $V$ and $V-I$ data 
from HST/WFPC2 if available, and the coordinates and the projected 
distance of each object from the center of NGC~3311.  From the foregoing 
discussion, we emphasize that these calculated masses should be viewed only as
plausible estimates and might be, if anything, lower limits.

\section{Discussion} 
\label{discuss}

The masses of these high-end GCs are all above $6 \times 10^6 M_{\odot}$ and extend to
almost $3 \times 10^7 M_{\odot}$. From their radial distribution, they are clearly within the
Hydra cluster (and possibly specifically associated with its cD galaxy, NGC~3311).
Structurally, they are very compact: A normal dwarf-galaxy scale length of 300 pc
\citep{deady02} would give $FWHM \sim 2\arcsec$ on our GMOS images, whereas our ``superluminous
GCs" are completely unresolved at the $0.5\arcsec$ resolution of our GMOS images, implying that
their scale radii are $r_{eff} \lesssim 50$ pc. A better limit on their scale sizes would,
however, come from HST imaging with its $0\farcs1$ resolution. Ten of our UCD candidates fall
within the WFPC2 field near NGC~3311 studied by \citet{brodie00} and two additional UCD
candidates fall on the WFPC2 archival data for NGC 3309. We extracted these images from the HST
Archive and measured them. We find that the 12 candidates, all on the undersampled WF frames,
have FWHMs ranging from $0\farcs204$ to $0\farcs250$, averaging $0\farcs222 \pm 0\farcs006$. By
comparison, the unresolved PSF on WF2,3,4 has a measured $\langle FWHM \rangle = 0\farcs208 \pm
0\farcs004$. The UCD candidates are therefore {\sl marginally} resolved (at the $2-\sigma$
level). Subtracting $FWHM_{UCD}$ in quadrature from $FWHM_{PSF}$, we can then estimate {\sl
very roughly} that our candidates have scale sizes equivalent to an effective diameter
of $\simeq20$ pc. This is only a crude estimate but is precisely in the $r_{eff}$ range occupied
by the known Fornax UCDs, most of the Virgo UCDs, and the DGTOs \citep{evsti07,hasegan05}. In
Table 1, we list the $V$ and $(V-I)$ measurements of the 12 overlapping candidates as obtained
from the WFPC2 data. They average $\langle V-I \rangle_0 = 1.1$, entirely similar to
normal red-sequence GCs.

There are now three main scenarios to explain UCDs. One possibility is that UCDs are the
nuclei of dE,N galaxies that have been stripped of their envelopes via galaxy ``threshing,''
on multiple passes through a larger galaxy \citep{bekki01}. \citet{fk02} suggest that UCDs
may also form from the agglomeration of young massive star clusters in locations of ongoing,
violent star formation. A third possibility is that UCDs are simply high-mass extensions of
globular clusters and share a common formation mechanism with their lower mass counterparts.
Evidence can be found for each of these formation scenarios \citep[e.g.][]{hasegan05},
suggesting that UCDs may not be a homogenous population; rather, objects can end up as UCDs
in different ways.

\citet{rejkuba07} and \citet{barmby07} provide strong new evidence from M/L ratios and
structural sizes of the most massive known GCs in NGC 5128 and M31 that they may form the
beginning of the long-missing bridge between the GC and dwarf-E sequence. It has long been
thought that GCs had a constant scale size $r_h \sim 3$ pc independent of mass, whereas $r_h
\sim M^{0.6}$ for E galaxies \citep[e.g.][]{hasegan05, barmby07}. This new evidence suggests
that massive star clusters ($10^7 M_{\odot}$ and above) must somehow form at increasingly
larger scale size regardless of their environment. Recent work by \citet{evsti07} on the
Fornax and Virgo UCDs in the range of $10^7 - 10^8 M_{\odot}$ further traces out a continuous
sequence between globular clusters, UCDs, dE nuclei and de,Ns, and giant ellipticals
in velocity dispersion and magnitude space. A clear sequence also exists in the
$\kappa_1-\kappa_3$ plane of $\kappa$-space, a fundamental plane for dynamically hot systems
originally defined by \citet{bender92}.

The UCD candidates in NGC~3311 mark out the clearest connection of such objects with GCs 
yet found within any one galaxy. This result is consistent with the 
idea that there may be a 
smooth bridge in structural parameters between these and the 
high-mass UCDs. Indeed, given that evidence has
been presented to support a) the idea that UCDs can form via diverse mechanisms, rather than
a single, evolutionary path \citep[e.g.][]{hasegan05,evsti07}, and b) the existence of a
continuous sequence between these objects in structural parameter space, it seems that
regardless of how a compact stellar system is assembled, it will strongly converge to a
structure that falls within this unified sequence. The sequence in NGC~3311 provides an excellent
opportunity to further trace this ``bridge" of intermediate-mass old stellar systems. A
logical next step would be radial velocity measurements, which would help decide whether they
belong more to NGC~3311 (and thus formed along with its GCs) or to the general Hydra cluster.

\acknowledgements

EHW and WEH would like to thank the Natural Sciences and Engineering Research 
Council of Canada (NSERC) for their funding of this project.  
The authors also thank Kyle Johnston for his help with world coordinate 
systems and the anonymous referee for his 
helpful comments.  This research has made use of the NASA/IPAC 
Extragalactic Database (NED) which is operated by the Jet Propulsion 
Laboratory, California Institute of Technology, under contract with the 
National Aeronautics and Space Administration.

\begin{deluxetable*}{cccccccccc}
\renewcommand\baselinestretch{1}
\tabletypesize{\small}
\tablehead{
\colhead{UCD}
& $M_{g'}$ & $Mass$ & $i'$ & $(g'-i')_0$ & $V$ & $V-I$ & $RA$ & $Dec$ & $D$\\
 & & ($10^6 M_{\odot}$) & & & & & & & ($Kpc$) 
}
\label{tab1}
\tablecaption{Properties of the UCD candidates}
\startdata

 1  &  -12.3  &   27.1  &   20.67   &   0.84  &    -    &    -   &  10:36:43.320    &   -27:29:24.19  &   36.1 \\
 2  &  -11.9  &   17.9  &   21.16   &   0.80  &    -    &    -   &  10:36:32.456    &   -27:29:24.88  &   51.0 \\
 3  &  -11.5  &   12.4  &   21.37   &   0.99  &    -    &    -   &  10:36:41.160    &   -27:31:21.90  &   7.6  \\
 4  &  -11.5  &   12.6  &   21.43   &   0.91  &  22.24  &  1.15  &  10:36:47.481    &   -27:31:10.59  &   18.3 \\
 5  &  -11.4  &   11.7  &   21.46   &   0.96  &    -    &    -   &  10:36:44.923    &   -27:34:20.21  &   42.6 \\
 6  &  -11.3  &   10.5  &   21.55   &   0.99  &  22.54  &  1.28  &  10:36:40.639    &   -27:32:06.42  &   10.1 \\
 7  &  -11.4  &   11.2  &   21.55   &   0.92  &    -    &    -   &  10:36:50.016    &   -27:31:57.35  &   25.7 \\
 8  &  -11.2  &   9.8   &   21.71   &   0.91  &    -    &    -   &  10:36:34.899    &   -27:29:44.90  &   41.3 \\
 9  &  -11.1  &   8.9   &   21.81   &   0.91  &    -    &    -   &  10:36:43.487    &   -27:30:26.04  &   19.9 \\
10  &  -11.0  &   9.2   &   21.81   &   0.87  &    -    &    -   &  10:36:49.500    &   -27:30:48.07  &   27.3 \\
11  &  -11.1  &   8.3   &   21.82   &   0.97  &    -    &    -   &  10:36:42.354    &   -27:31:00.00  &   10.9 \\
12  &  -11.0  &   8.2   &   21.87   &   0.94  &    -    &    -   &  10:36:41.249    &   -27:31:07.80  &   10.3 \\
13  &  -11.3  &   10.4  &   21.87   &   0.68  &    -    &    -	 &  10:36:43.186    &   -27:29:59.72  &   26.8 \\
14  &  -10.9  &   7.5   &   21.89   &   1.01  &  22.84  &  1.27  &  10:36:43.578    &   -27:32:17.43  &   10.0 \\
15  &  -11.0  &   8.1   &   21.90   &   0.92  &  22.81  &  1.18  &  10:36:48.133    &   -27:32:26.02  &   22.1 \\
16  &  -11.0  &   8.2   &   21.92   &   0.89  &  23.03  &  1.22  &  10:36:36.203    &   -27:32:19.59  &   25.2 \\
17  &  -11.0  &   7.9   &   21.94   &   0.91  &    -    &    -   &  10:36:40.978    &   -27:30:20.10  &   22.3 \\
18  &  -11.0  &   8.2   &   21.97   &   0.84  &    -    &    -   &  10:36:41.290    &   -27:31:35.55  &   5.5  \\
19  &  -11.0  &   8.0   &   21.98   &   0.86  &  22.65  &  1.07  &  10:36:48.321    &   -27:30:42.38  &   24.8 \\
20  &  -11.1  &   8.3   &   22.00   &   0.79  &  22.78  &  1.12  &  10:36:45.606    &   -27:31:30.90  &   10.2 \\
21  &  -10.8  &   6.7   &   22.01   &   1.02  &    -    &    -   &  10:36:42.100    &   -27:31:05.70  &   9.6  \\
22  &  -10.9  &   7.3   &   22.05   &   0.89  &  22.96  &  1.22  &  10:36:45.055    &   -27:31:53.86  &   8.6  \\
23  &  -10.9  &   7.2   &   22.09   &   0.86  &  23.05  &  1.23  &  10:36:50.646    &   -27:32:02.50  &   28.1 \\
24  &  -10.8  &   6.9   &   22.10   &   0.89  &    -    &    -   &  10:36:47.876    &   -27:29:19.14  &   41.4 \\
25  &  -10.9  &   7.2   &   22.10   &   0.85  &    -    &    -   &  10:36:41.556    &   -27:31:23.32  &   6.4  \\
26  &  -10.7  &   6.2   &   22.10   &   1.01  &    -    &    -   &  10:36:39.316    &   -27:30:53.73  &   17.4 \\
27  &  -10.8  &   6.6   &   22.10   &   0.94  &  23.04  &  1.16  &  10:36:44.577    &   -27:30:48.51  &   15.2 \\
28  &  -10.7  &   6.2   &   22.11   &   1.01  &  23.24  &  1.33  &  10:36:39.024    &   -27:31:22.96  &   14.0 \\
29  &  -10.8  &   6.7   &   22.15   &   0.88  &  23.05  &  1.16  &  10:36:48.803    &   -27:31:53.72  &   21.3 \\
                                                                   
\enddata

\end{deluxetable*}


%
%


%
%


%
%

\end{document}